\documentclass[11pt]{advances2}

\usepackage[english]{babel}

\usepackage{amssymb}
\usepackage{latexsym}
\usepackage{graphicx}
\usepackage{xspace}

\usepackage{pstricks,pst-node}

\newtheorem{theorem}{Theorem}[section]

\newtheorem{proposition}[theorem]{Proposition}

\newtheorem{coro}[theorem]{Corollary}
\newtheorem{theo}[theorem]{Theorem}

\theoremstyle{definition}

\theoremstyle{remark}
\newtheorem{remark}[theorem]{Remark}

\numberwithin{equation}{section}

\copyrightinfo{}
{}

\newcommand{\R}{{\mathbb R}}
\newcommand{\LL}{{\mathbb L}}

\newcommand{\CC}{C_{p,q}}

\begin{document}
\setcounter{page}{1}

\title{Recent progress on the notion of global hyperbolicity}

\author[S\'anchez]{Miguel S\'anchez}
\address{Departamento de Geometr\'{\i}a y Topolog\'{\i}a \\ Universidad de Granada\\ Facultad de Ciencias, Campus de Fuentenueva s/n E-18071 Granada, Spain}
\email{sanchezm@ugr.es}


\date{\today}
\maketitle

\begin{abstract}
Global hyperbolicity is a central concept in Mathematical
Relativity. Here, we review the different approaches to this
concept explaining both, classical approaches and recent results.
The former includes Cauchy hypersurfaces, naked singularities, and
the space of the causal curves connecting two events. The latter
includes structural results on globally hyperbolic spacetimes,
their embeddability in Lorentz-Minkowski, and the recently revised
notions of both causal and conformal boundaries. 
Moreover, two
criteria for checking global hyperbolicity are reviewed. The first
one applies to general splitting spacetimes. The second one
characterizes accurately global hyperbolicity and spacelike Cauchy
hypersurfaces for standard stationary spacetimes, in terms of a
naturally associated Finsler metric.
\end{abstract}

\section{Introduction and Notation}

Global hyperbolicity is a central concept in Mathematical
Relativity, which is involved in almost all global questions in
this area, since the initial value problem to cosmic censorship
(see for example \cite{Sa-bilbao}). The notion was introduced by
Leray \cite{Le} in 1953, and developed in the {\em Golden Age of
General Relativity} by Avez, Carter, Choquet-Bruhat, Clarke,
Hawking, Geroch, Penrose, Seifert and others. However, some
questions which affected basic approaches to this concept,
remained unsolved in this epoch.

Concretely, the so-called {\em folk problems of smoothability}
\cite{SW}, affected the differentiable and metric structure of any
globally hyperbolic spacetime $M$. Their ramifications include the
possible embeddability of $M$ in some Lorentz-Minkowski space
$\LL^N$ (in the spirit of Nash theorem), and other issues in the
consistency of the {\em causal ladder} of spacetimes. Moreover,
the (GKP) {\em causal boundary}  \cite{GKP} introduced a new
ingredient for the causal structure of spacetimes, as well as a
new viewpoint for global hyperbolicity. However, the lack of a
full consistency for this boundary (especially in relation with
the {\em conformal boundary}), remained as an open issue since
this epoch.

Recently, these old issues have been revisited, and a full
solution seems available now. Our purpose in this note is to give
a brief account of both, the old issues and the recent progress,
focused on the notion of global hyperbolicity.

More precisely, in Section \ref{s2}, the alternative definitions
of global hyperbolicity in terms of topological elements of the
spacetime (Cauchy hypersurfaces, absence of naked singularities)
are explained. The equivalences  are based in the central article
by Geroch \cite{Ge}, and include a recent conceptual
simplification in \cite{BeSa4} (see Theorem \ref{t2}). Section
\ref{s3} is devoted to the implications of the  folk problems of
smoothability on the differentiable and metric structure of
globally hyperbolic spacetimes (Theorem \ref{t3}). Such a
structure yields also a characterization of these spacetimes in
terms of their embeddability in Lorentz-Minkowski (Proposition
\ref{p3}). In Section \ref{s4} original Leray's definition of
global hyperbolicity is considered. This is expressed in terms of
the space of (continuous) causal curves connecting two events.
Such a space present some subtleties which are pointed out. In
Section \ref{s5}, the problem and recent solution to the notion of
causal boundary is briefly explained. Then, a new characterization
of global hyperbolicity is stated (Theorem \ref{t5}). Moreover,
the conditions under which the conformal boundary characterizes
global hyperbolicity are also enounced (Theorem \ref{t5b}). The
section is ended with a scheme about the different equivalent
approaches to global hyperbolicity.

A different question is to determine, for a concrete spacetime,
its possible global hyperbolicity. In the last two sections this
question is studied for two  families of spacetimes. The first one
(Section \ref{s6}), is the class of spacetimes which admit a
smooth splitting type $\R\times S$ and such that the natural
coordinate $t\in\R$ is a temporal function, being its levels
$\{t_0\}\times S$ the natural candidates for Cauchy hypersurfaces
(Theorem \ref{t6}). The second one (Section \ref{s7}), is the
subclass of previous spacetimes $M=\R\times S$ which contains all
standard stationary spacetimes. In these spacetimes, one
characterizes exactly both, when the spacetime is globally
hyperbolic and, in this case, when the  slices $\{t_0\}\times S$
are Cauchy hypersurfaces. Such a characterization is expressed in
terms of an auxiliary Finsler metric on $S$.

 Throughout this paper, we will use
standard notation in Causality, as in the books and reviews
\cite{BEE, GPS, HE, MiSa, O, PeLibro, W}. In particular, a
spacetime will be a connected time-oriented $n$-manifold ($n\geq
2$), its chronological and causal relations are denoted $\ll,
\leq$, resp., and these relations determine the chronological and
causal futures and pasts $I^\pm(A), J^\pm(A)$ of any point or
subset $A$ of $M$. We also put:
$$
\begin{array}{c}
I(p,q):= I^+(p)\cap I^-(q),, \quad J(p,q):= J^+(p)\cap J^-(q), \\
\uparrow P:= I^+(\{x\in M: p\ll x, \; \forall p\in P\})
\\
\downarrow F:=I^-(\{x\in M: x\ll q, \; \forall q\in F\})
\end{array}$$ for any $p,q\in M$ and $ P,F \subset M$. Timelike
and causal curves are regarded as smooth ($C^1$ is enough) curves
with timelike or causal derivative, except in Section \ref{s4},
where {\em continuous causal} curves are explicitly considered.

\section{Topological equivalences on the manifold}\label{s2}
The simplest way to understand global hyperbolicity relies on the
interplay between the causality of the spacetime and some
topological elements of the manifold. Concretely, recall the
following ones:

\begin{itemize}
\item {\em Cauchy hypersurface:} subset $S\subset M$ which is
crossed exactly once by any inextensible timelike curve.

Then, $S$ must be an embedded topological hypersurface and must be
also crossed by any inextensible causal curve $\gamma$ \cite{Ga2,
O}. However, such a $\gamma$ may intersect $S$ not only in a point
but also along a compact interval of its domain. If causal curves
cannot intersect  $S$ in more than one point, then $S$ is an {\em
acausal Cauchy hypersurface}.

Easily, the existence of a Cauchy hypersurface $S$ implies that
$M$ is homeomorphic to $\R\times S$, and all Cauchy hypersurfaces
are homeomorphic.

\item {\em Time function:} continuous function $t: M\rightarrow
\R$ which increases strictly on any future-directed causal curve.

If, moreover, the levels $t=$constant are Cauchy hypersurfaces,
then $t$ is a {\em Cauchy time function}. By convenience, all
Cauchy functions will be assumed onto (this is not restrictive
because, otherwise, a re-scaling of $t$ will fulfill this
property).

\item {\em Absence of  naked singularities}: this means  that
$J(p,q)$ 
is compact for all $p, q$ in $M$.

In fact, if $J(p,q)$ is non-compact for some $p, q$, then one can
check the  existence of a future-directed causal curve $\rho$
starting at $p$ and contained in $J(p,q)$, with no endpoint in the
closure $\overline{J(p,q)}$. Moreover, this curve is always
visible from $q$, i.e., any point $\rho(s)$ can be joined with $q$
by means of a future-directed causal curve. Summing up, in this
case one can say that a (physical) {\em naked singularity} appears
between $p$ and $q$ (see also \cite{Clb}).
\end{itemize}
The connections among these elements are summarized in the
following result.
\begin{theorem}\label{t2}
For a spacetime $M$, the following items are equivalent:
\begin{enumerate}
\item $M$ is causal and does not have naked singularities.

\item $M$ is strongly causal and does not have naked
singularities.

\item $M$ admits a Cauchy time function $t$.

 \item $M$ admits a Cauchy hypersurface $S$.

\end{enumerate}
$M$ is called globally hyperbolic when such equivalent items
holds.
\end{theorem}
The following comments on Theorem \ref{t2} are in order. Item (2)
is a typical definition of global hyperbolicity, which is used in
standard books such as \cite{BEE, HE, O, PeLibro, W}. Item (1) is
an even simpler definition in \cite{BeSa4}. In fact, (1)
$\Rightarrow$ (2) because the assumptions (1) imply that $M$ is
{\em causally simple} (as $J^\pm(p)$ is closed for all $p$) and,
therefore, strongly causal, as required. It is worth pointing out
that, under strong causality, the compactness of the closures
$\overline{J(p,q)}$ is enough to ensure global hyperbolicity
\cite[Lemma 4.29]{BEE}, but under causality such a property is not
enough  (Carter's example \cite[p. 195]{HE}, see \cite{Ga}).

The implication (2) $\Rightarrow$ (3) relies on the following
celebrated idea by Geroch \cite{Ge} (this author use the
definition of global hyperbolicity explained in Section \ref{s4},
so, see also \cite{Sa-bras}). First, one takes any finite measure
$m$ on $M$ associated to some Riemannian metric --or, with more
generality, one can take also any  {\em admissible measure} in the
sense of Dieckmann \cite{Di}). Then, define the past and future
volume functions $t^\pm: M\rightarrow \R$, as $t^-(p):= m(I^-(p)),
t^+(p):= -m(I^+(p))$. These functions are time functions for {\em
causally continuous} spacetimes (a class of spacetimes more
general than the causally simple ones and, so, which includes all
the globally hyperbolic spacetimes). Moreover, if $\gamma:
(a,b)\rightarrow M$ is an inextensible future-directed causal
curve, one has:
$$\lim_{s\rightarrow b} t^+(\gamma(s)) =0 =
\lim_{s\rightarrow a} t^-(\gamma(s)) \quad \lim_{s\rightarrow a}
(-t^+(\gamma(s))) , \lim_{s\rightarrow b} t^-(\gamma(s)) >0$$ so
that $t(z)= {\rm log} \left(-t^-(z)/t^+(z)\right)$ satisfies:
$$ \left.
\begin{array}{l}
\lim_{s\rightarrow b} t(\gamma(s)) = \infty \\
\lim_{s\rightarrow a} t(\gamma(s)) = -\infty
\end{array}
\right\} \Longrightarrow \hbox{levels $t=$const. are Cauchy}
$$
i.e., $t$ is the required Cauchy time function.

As the implication (3)$\Rightarrow$ (4) is trivial, all the
equivalences hold if one proves (4)$\Rightarrow$ (1). This follows
by using nowadays standard arguments on limit curves \cite{BEE} or
quasilimits \cite{O}. In fact, the absence of naked singularities
follows  from the compactness of the sets type $J^-(q)\cap J^+(S),
q\in M$ (and to prove causality is now an exercise simpler than
strong causality), see  \cite{BEE, Ga2, O}.

\section{Smoothability and structural results}\label{s3}

In previous section, the involved elements were defined at a
topological level, and we was not worried about its
differentiability. But, of course, this property will turn out
essential for applications. So, we will say that a Cauchy
hypersurface or time function is {\em smooth} if it is as
differentiable as allowed by the order of differentiability of the
spacetime. However, smoothness will not be enough for relevant
applications and, so, we state the following notions:

\begin{itemize}
\item {\em Spacelike Cauchy hypersurface:} a smooth Cauchy
hypersurface $S$ such that all the tangent hyperplanes $T_pS$,
$p\in S$, are spacelike.

Necessarily, $S$ is then acausal, but notice that a smooth acausal
Cauchy hypersurface may be non-spacelike, as the tangent
hyperplanes may be degenerate. From a technical viewpoint, the
initial value problem starts typically with a smooth Riemannian
manifold which will be a posteriori a spacelike Cauchy
hypersurface of the evolved spacetime.

Easily, the existence of a smooth Cauchy hypersurface $S$ implies
that $M$ is diffeomorphic to $\R\times S$, and all smooth Cauchy
hypersurfaces are diffeomorphic.

\item {\em Temporal function:} smooth time function $t:
M\rightarrow \R$ such that its gradient $\nabla t$ is timelike
--or, equivalently, a smooth function $t$ with past-directed
timelike
 gradient.
If, additionally, the levels $t=$constant are Cauchy hypersurfaces
(necessarily spacelike ones), then $t$ is a {\em Cauchy temporal
function}, and, again, $t$ will be assumed onto with no loss of
generality.

The existence of a Cauchy temporal function $t$ is especially
interesting, because it is equivalent to the existence of a global
orthogonal  splitting $M\equiv (\R\times S,g)$, where $g$ can be
written as
\begin{equation}\label{e1}
 g=-\beta dt^2 + g_t ,
\end{equation}
being $\beta$ (the {\em lapse}) a function on $\R\times S$, and
(under  a natural identification) $g_t$ a Riemannian metric on
each slice $\{t\}\times S$ varying smoothly with $t\in \R$.

\item {\em Steep temporal function}: a temporal function $t$ whose
gradient satisfies $|\nabla \tau|^2 (:= -g(\nabla t, \nabla
t))\geq 1$.

The existence of a steep temporal function is interesting for a
spacetime $M$, as it solves the problem of its isometric
embeddability in some $\LL^N$ (in the spirit of Nash' theorem
\cite{Na}), see \cite{MuSa}. In fact, it was noticed by Greene
\cite{Gr} and Clarke \cite{Cl} that any semi-Riemannian (or even
degenerate) manifold can be isometrically immersed in some
semi-Euclidean space $\R^N_s$ of sufficiently big dimension $N$
and index $s$. The problem is a bit subtler for $s=1$, but a
simple argument in \cite{MuSa} shows that a spacetime can be
isometrically embedded in $\LL^N$ if and only if it admits a steep
temporal function.

Moreover, a spacetime which admits a {\em steep Cauchy temporal
function} can be split as in (\ref{e1}) with a bounded lapse
function, as $\beta= |\nabla \tau|^{-2}$.

\end{itemize}
The equivalence of previous elements with global hyperbolicity can
be summarized as follows.

\begin{theorem}\label{t3}
For a spacetime $M$, any of the following items is equivalent to
global hyperbolicity:
\begin{enumerate}
\item  $M$ admits a  spacelike Cauchy hypersurface.

\item $M$ admits a Cauchy temporal function $t$ or, equivalently,
$M$ admits a splitting $M\equiv (\R\times S, g)$ with $g$ as in
(\ref{e1}).

 \item $M$ admits a steep Cauchy  temporal function or, equivalently, a
 global splitting $M\equiv (\R\times S, g)$ where $g$ adopts the form (\ref{e1}) with  $\beta<1$.

\end{enumerate}
\end{theorem}
The following comments on this theorem are in order. The so-called
{\em folk problems of smoothability}  consist in the question
whether the alternative definitions of global hyperbolicity in
Theorem \ref{t2} imply the items (1) or (2) in Theorem \ref{t3}
(or, with more generality, if certain type of causally constructed
continuous elements can be obtained also in a smooth way,  with an
appropriate causal character). The first problem (glob. hyp.
$\Rightarrow$ Theorem \ref{t2}(1)) was posed explicitly in
\cite[p. 1155]{SW} and solved in \cite{BeSa1}. The second one
(glob. hyp. $\Rightarrow$ Theorem \ref{t2}(2)) was solved in
\cite{BeSa2}. Moreover, here  the question whether any spacetime
which admits a time function must admit a temporal function too,
is answered affirmatively. This question affected the consistency
of the two classical definitions of {\em stable causality} and,
thus, the structure of the so-called {\em causal hierarchy of
spacetimes} (see the review \cite{MiSa} for full details).

The difficulty in the solution of these problems relied in two
facts:
\begin{itemize}
\item[(a)] Notice that the volume functions $t^\pm$ in the proof
of Geroch's theorem are continuous in the globally hyperbolic
case, and there exists a big freedom of admissible measures in
order to construct them. Nevertheless, in general one cannot
expect that, for example, if $t^\pm$ is only a continuous time
function constructed from some admissible measure then, by
changing this measure, the new functions $t^\pm$ will be smooth.
In fact, $t^\pm$ are time functions if and only if the spacetime
is causally continuous \cite{MuSa}, that is, $t^\pm$ are not
continuous if the spacetime is only stably causal. Nevertheless,
in such a spacetime, the existence of a time function is ensured
thanks to an original argument by Hawking \cite{Ha2}. Of course,
to change the admissible measure is useless for this argument, as
$t^\pm$ are always non-continuous.

\item[(b)] The smoothability problems not only affect
smoothability, but also the causal character of the involved
elements. If, for example, some sort of smoothing procedure (say,
some type of convolution) would yield a smooth Cauchy hypersurface
$S$, one would have to study still the case when $S$ is
degenerate. And this would be delicate, as a small perturbation of
$S$ (in order to get a spacelike hypersurface) might spoil the
Cauchy character.
\end{itemize}
Due to these two difficulties, the proofs in \cite{BeSa1, BeSa2}
are based in the construction of some sort of semi-local temporal
functions and a systematic process of sum, very different to the
arguments in the proof of Theorem \ref{t2}. However, the Cauchy
temporal function constructed by Geroch \cite{Ge}, is required in
order to start the process.

It is also worth pointing out that the Cauchy temporal function
$t$ can be chosen such that any prescribed spacelike hypersurface
$S$ is one of its levels (and $S$ can be chosen such that any
prescribed acausal compact spacelike submanifold with boundary is
included in $S$), \cite{BeSa3}. This contributes to the
consistency of the usual procedures in Mathematical
Relativity\footnote{For example, in the initial value problem, one
could conceive the following situation: the initial hypersurface
is a Cauchy hypersurface of the evolved spacetime, but it is not a
slice of any Cauchy temporal function. In this case,  well-posed
initial data on $S$ would imply structural restrictions in the
evolved spacetime.}.

The item (3) becomes relevant in both ways: it is a refinement of
the structural decomposition (\ref{e1}) for any globally
hyperbolic spacetime, and it implies the isometric embeddability
of all globally hyperbolic spacetimes in Lorentz-Minkowski. This
embeddability had been already claimed by Clarke \cite{Cl};
however, his proof was affected by the folk problems of
smoothability. The proof in \cite{MuSa} is based in a constructive
procedure of a steep Cauchy temporal function,  which also starts
at Geroch's Cauchy time function. This construction is independent
and easier than the one in \cite{BeSa2}. However, it is carried
out specifically for globally hyperbolic spacetimes. Thus, in
principle, it cannot be used to construct a temporal function in
any stably causal spacetime.

Finally, it is worth pointing out that, for a spacetime which
admits a temporal function but it is not globally hyperbolic, the
existence of a steep temporal function can be lost or gained by
changing conformally the metric \cite{MuSa}. So, one has:

\begin{proposition}\label{p3} Let $M$ be a spacetime. Then:

(A) $M$ is globally hyperbolic if and only if all the spacetimes
in its conformal class are isometrically embeddible in some
Lorentz-Minkowski space $\LL^N$ of sufficiently high dimension
$N$.

(B) $M$ is stably causal (i.e., $M$ admits a temporal function) if
and only if some representative of its conformal class can be
isometrically embedded in some Lorentz-Minkowski space $\LL^N$.
\end{proposition}

\section{The space of continuous causal curves}\label{s4}

Historically, the notion of global hyperbolicity was introduced by
Leray \cite{Le}  starting at the the space of causal curves $\CC$
connecting two points, in a wider frame for hyperbolic equations.
This first notion was developed fast (for example, see \cite{Av,
CB, Ha, Pe, Se}). It is worth pointing out some technicalities on
this space.

First, all these curves will be taken reparameterized in the same
interval, namely $I=[0,1]$, so that the compact-open topology will
be assumed in $\CC$. However, sequences of (smooth) causal curves
will have non--smooth limits in a natural way, and these limits
must be regarded as causal too. So, a {\em (future-directed)
continuous causal curve} $\gamma: I\rightarrow M $ is defined as a
(continuous) curve which, for each convex
neighbourhood\footnote{i.e., $U$ is a (starshaped) normal
neighbourhood of all its points (see \cite[p. 129]{O}).} $U\subset
M$, satisfies: if $t,t' \in I, t \leq  t'$ with $\gamma([t,t'])
\subset U$, then $\gamma(t)\leq_U \gamma(t')$ (where $\leq_U$
denotes the causal relation in $U$, regarded as a spacetime). One
can check that, for a (continuous) curve $\gamma$ defined on $I$
and non locally constant around any point $t_0\in I$, the
following equivalence holds \cite[Appendix B]{CFS}: $\gamma$ is
continuous causal iff $\gamma$ is $H^1$ and $\dot \gamma(s)$ is a
future--directed causal vector for almost all $s\in I$ --in
particular, $\gamma$ is Lipschitzian\footnote{In this Lorentzian
setting, concepts such as $H^1$ or Lipschitzian (or uniform
convergence below) can be regarded as those for any auxiliary
Riemannian metric $h$.}. The space $\CC$ is then the set of all
these continuous causal curves  which connect $p$ with $q$,
endowed with the compact-open topology (or equivalently, with the
topology of uniform convergence).

However, one has still the problem that all the
reparameterizations of a single curve $\gamma\in \CC$ yields a
non-compact subset. Following Choquet-Bruhat \cite{CB}, one can
fix some auxiliary Riemannian metric $h$ and consider just
 the subset:
 $$
 \CC^h =\{\gamma\in \CC: h(\dot\gamma,\dot\gamma)=c \;
 \mbox{(constant)}\; \hbox{a.e.}\}.
$$
We emphasize that, even in $\LL^N$, the space  $\CC^h$ is not
compact whenever $p\ll q$ (this, as well as other subtleties along
this section, have been studied in \cite{BeSaUnp}). However, the
closure of $\CC^h$ in $\CC$ will be compact, which will be enough
for our purposes.

\begin{theorem}
For a  spacetime $M$, the following items are equivalent:
\begin{enumerate}
\item  $M$ is globally hyperbolic.
 \item For each $p,q\in M$, the
set  of all the continuous causal  curves $\CC^h$ (parametrized on
$[0,1]$ at constant speed for an auxiliary Riemannian metric $h$)
has a compact closure in the space of all the continuous causal
curves $\CC$ endowed with the compact-open topology.
\end{enumerate}
\end{theorem}
In fact, (2) $\Rightarrow$ (1)  is now straightforward, as one can
check easily that (2) implies both the absence of naked
singularities and causality (remarkably, the reasoning by
Choquet-Bruhat in \cite{CB} proved directly strong causality). The
converse can be proved by using known tools of limit curves.

Many of the bothering subtleties for $\CC$ come from the
reparameterizations of the curves. In order to avoid this, we will
call the image of a continuous causal curve a (causal) {\em path}.
Notice that, if a causal curve $\alpha$ is closed, then going two
rounds along it we obtain a new curve $\alpha^2$. Obviously,
$\alpha$ and $\alpha^2$ yield the same path, but none of them is a
reparameterization of the other one. Such paths will be excluded
by considering causal spacetimes. In these spacetimes, a path can
be regarded also as the class of a continuous causal curve in
$\CC^h$ up to a reparameterization. A natural topology for paths
is the $C^0$ one --namely, a path $\rho$ is the limit of a
sequence $\{\rho_n\}_n$ if any open set $U\subset M$ which
contains $\rho$ contains also all but a finite number of $\rho_n$,
see \cite{BEE}. Good properties on convergence for the $C^0$
topology appear in strongly causal spacetimes. However, as the
endpoints of the paths in $\CC$ are fixed, causality will be
enough for the following characterization.

\begin{theorem}
For a causal spacetime $M$, the following  are equivalent:
\begin{enumerate}
\item  $M$ is globally hyperbolic.

\item For each $p,q\in M$, the space of continuous causal paths
$\CC^{path}$ (i.e., the set of the images of continuous causal
curves in $\CC$) endowed with the $C^0$ topology, is compact.
\end{enumerate}
\end{theorem}
In fact, (2) $\Rightarrow$ (1)  can be found in Geroch's
\cite{Ge}, and the converse follows from known properties of the
$C^0$ convergence which goes back to \cite{Se}.

Finally, it is worth pointing out that, because of the compactness
properties of $\CC$, and the lower  continuity of the energy
functional on $\CC$, one can prove easily the following
Avez-Seifert property \cite{Av, Se}: {\em if two points $p\neq q$
of a globally hyperbolic spacetime $M$ are causally related
($p<q$), then they can be connected by means of a causal geodesic
with length equal to the time-separation (Lorentzian distance)
$d(p,q)$}. In particular, $d(p,q)$ is always finite in globally
hyperbolic spacetimes. However, for non-globally hyperbolic
spacetimes, the finiteness of $d(p,q)$ at some $p,q$, will be lost
for some representatives of the conformal class \cite[Th.
4.30]{BEE}. Summing up:

\begin{proposition} For a strongly causal  spacetime $M$, the following properties are equivalent:

(1) $M$ is globally hyperbolic.

(2) The Lorentzian distance $d$ is finite valued for all the
spacetimes in the conformal class of $M$.
\end{proposition}

\section{Causal and conformal boundaries}\label{s5}
Among the notions of boundary for a spacetime, the conformal and
causal ones are the most useful and promising in Mathematical
Relativity. Global hyperbolicity is closely related to the
properties of these boundaries; in fact, it is commonly claimed
that a spacetime is globally hyperbolic when the causal or
conformal boundary does not contain a {\em timelike point}.
Nevertheless, both boundaries have presented problems of
consistency, which only recently have been solved. So, we
summarize very briefly these problems and how global hyperbolicity
can be characterized in terms of these boundaries. We refer to
\cite{FHS} for exhaustive discussions and references.

The notion of causal boundary $\partial_cM$ for a spacetime $M$
was introduced by Geroch, Kronheimer and Penrose \cite{GKP} ({\em
GKP boundary}). The idea was to attach a boundary $\partial_cM$ to
to any strongly causal spacetime so that each inextensible
future-directed (resp. past-directed) timelike curve will have an
endpoint in $\partial_cM$. The initial idea was that two such
future-directed (resp. past-directed) timelike curves $\gamma,
\tilde\gamma$ must reach the same point when their chronological
pasts (resp. future) coincide, i.e.,
$I^-(\gamma)=I^-(\tilde\gamma)$ (resp.
$I^+(\gamma)=I^+(\tilde\gamma)$). So all these past (resp. future)
sets or {\em TIPs} (resp. {\em TIFs}) would be regarded as
boundary points. However, two problems appear. The first one is
that, sometimes, it is natural to expect that both, an
inextensible future-direct timelike curve and a past-directed one,
will have the same endpoint, i.e., a TIP and a TIF would be
identified as the same boundary point\footnote{Think, for example,
in $M=\{(x,t): x>0\}\subset \LL^2$. Each $(0,t)\in \LL^2$ yields
naturally a TIP, $P=I^-((0,t))\cap M$, and a TIF,
$F=I^+((0,t))\cap M$, which should be regarded as a unique
boundary point.}.
 The second one is to
topologize the completion --so that one can check precisely when a
sequence or curve in $M$ converges to a point in $\partial_cM$.
These two problems are closely related, and yielded a hard {\em
identification problem}, studied by many authors shortly after the
seminal GKP paper (see, for example, \cite{BS,  KLPRD, Sz}). The
main difficulty for this problem was that, apparently, there were
many possible choices of both, identifications and topologies.
Nevertheless, no one of them seemed to be naturally consistent in
the following sense: if $M$ is a simple open subset of
Lorentz-Minkowski $\LL^N$, $\partial_cM$ must agree with the
topological boundary of $M$ in $\LL^N$ --or, at least, a
satisfactory reason which justifies the discrepancy must be
provided.

A critical review on the different attempts to solve this problem
can be found in \cite{S}. Here, we point out just the following
elements of the solution provided in \cite{FHS}, which takes into
account the recent progress in \cite{F, H1, MR}. The causal
boundary $\partial_cM$ is composed by {\em timelike} points and
{\em non-timelike} points. The former are the pairs type $(P,F)$
where
 $P$ is a TIP, $F$ is a TIF and they are S-related, i.e., $P$ is
included and is maximal\footnote{Maximal in the sense that no
other TIP $P'$ satisfies $P\varsubsetneq P'\subset \downarrow F$.}
in the common past $\downarrow F$ of $F$ and, viceversa, $F$ is
included and is maximal in the common future $\uparrow P$. The
non-timelike points are pairs type $(P,\emptyset)$ or $(\emptyset,
F)$, where $P$ (resp. $F$) is a TIP (resp. TIF) which is not
S-related with any TIF (resp, TIP). The spacetime itself is also
regarded as the set of all the pairs $M\equiv \{(I^+(p),I^-(p)):
p\in M\}$ so that one has naturally the causal completion
$\overline M=M\cup
\partial_cM$, composed by pairs of subsets of $M$. The chronological relation $\ll$ is extended to a
natural chronology $\overline{\ll}$ in $\overline M$, namely:
$(P,F)\overline \ll (P',F')$ if $F\cap P'\neq \emptyset$.
Moreover, there exists a natural (but subtle) way to topologize
$\overline M$.

So, if $(P,F)\in \partial_cM$ is a timelike point, one has
$p\overline{\ll} (P,F) \overline{\ll} q$ for any  $p\in P$, $q\in
F$. Nevertheless, for a non-timelike point, say, $(P,\emptyset)$,
there is no other  $(P',F')\in \overline M$  which lies in its
$\overline \ll$--chronological future. It is not difficult to
check that the existence of a timelike point is equivalent to the
existence of a naked singularity (see \cite{FHS} for exhaustive
details). So, one has:

\begin{theorem}\label{t5}
For a strongly causal spacetime\footnote{An important reason to
impose strong causality is to ensure that the completion
$\overline M$ will have a natural satisfactory topology --namely,
the topology on $M$ will be the Alexandrov one, generated by the
sets type $I(p,q)$, and this topology coincides with the manifold
topology only in strongly causal spacetimes. In principle, the
notion of causal boundary as a point set endowed with a extended
chronological relation would make sense for causal spacetimes.
Harris \cite{H1} considered even a more general notion of
chronological set. However, some subtleties appear  when one likes
to ensure that, for any inextensible future-directed timelike
curve $\gamma$, the set  $I^-(\gamma)$ is truly a TIP, in the
sense of {\em terminal indecomposable set} introduced in
\cite{GKP}.} $M$, the following properties are equivalent:
\begin{enumerate}
\item  $M$ is globally hyperbolic.

\item The causal boundary $\partial_c M$ of $M$ does not admit a
timelike point, i.e., any pair $(P,F)\in \partial_c M$ satisfies
either $P= \emptyset$ or $F=\emptyset$.
\end{enumerate}
\end{theorem}

About the conformal boundary, the following  remarks are in order.
First, it is the most commonly used boundary in Mathematical
Relativity: the Penrose-Carter diagrams (or typical concepts such
as asymptotic flatness \cite{AH, W}) are stated  in terms of a
conformal boundary, see \cite{HE}. However, this boundary is just
an {\em ad hoc} construction for some classes of spacetimes. In
fact, it is defined by using an open conformal embedding
$i:M\hookrightarrow M_0$ and taking the topological boundary of
the image $\partial_iM := \partial (i(M))$. However, there is no a
general recipe which says when two such embeddings $i, j$ will
yield  conformal boundaries $\partial_iM, \partial_jM$ which are
isomorphic in some natural sense --or when a given spacetime will
admit some useful open conformal embedding.

As emphasized in \cite{FHS}, the best we can say is that, under
some hypotheses, the conformal boundary $\partial_iM$ (or some
part of it) will agree with the causal one $\partial_cM$. Then,
$\partial_iM$ will reflect intrinsic properties of $M$. In this
case, $\partial_iM$ may be very useful from a practical viewpoint,
as it may be much easier to compute than $\partial_cM$.

A first condition for the identification of both boundaries is the
following one. The open conformal embedding $i:M\hookrightarrow
M_0$ must be {\em chronologically complete}, i.e, if $\gamma$ is a
future-directed inextensible curve in $M$ then the curve
$i\circ\gamma$ (which is necessarily future-directed and timelike
in $M$) must have a future endpoint in $\partial(i(M))$ --and
analogously for past-directed curves. We will not go through more
details on the conditions for the identifications of both
boundaries, which are detailed in \cite{FHS}. Simply, we recall
the following characterization of global hyperbolicity, when one
can say that the conformal boundary is  $C^1$:
\begin{theorem}\label{t5b}
Let $M$ be a  spacetime which admits an open conformal embedding
$i:M \hookrightarrow M_0$ in a strongly causal spacetime $M_0$
such that:  (i) $i(M)$ is an open subset of $M_0$ with $C^1$
boundary, and (ii) $i$ is chronologically complete. Then, the
following properties are equivalent:
\begin{enumerate}
\item  $M$ is globally hyperbolic.

\item The conformal boundary $\partial_iM:=\partial (i(M))$ of $M$
does not admit a {\em timelike point}, i.e., no $z\in \partial_iM
(\subset T_zM_0)$ admits as tangent space $T_z(\partial_iM)$ an
hyperplane with Lorentzian signature.
\end{enumerate}
Moreover, in this case the causal boundary $\partial_cM$ is
naturally identified with the conformal one $\partial_iM$.
\end{theorem}
We remark that, in this theorem, the abstract notion of timelike
point for the causal boundary, is translated in the more palpable
notion of a point in the conformal boundary with timelike tangent
hyperplane. It is very easy to prove that (1) $\Rightarrow$ (2)
(in fact, the negation of (2) at a point $z$, yields directly a
naked singularity, and only the local properties around $z$ are
relevant for this). However, all the properties of the embedding
must be carefully used in order to obtain the reversed
implication, see \cite{FHS}.

The equivalences between all previous approaches to global
hyperbolicity are summarized in the figure in the next page.


\section{Checking global hyperbolicity in general splitting spacetimes}\label{s6}

Taken into account the Cauchy splitting (\ref{e1}), we can wonder,
conversely, when a spacetime splitted as in (\ref{e1}) admits as
Cauchy hypersurfaces the levels $t\equiv $ constant. Even more,
the applicability will be bigger if we admit also mixed terms
between the parts in $\R$ and $S$. This was studied systematically
in \cite{Sa-bari}, and the results are summarized next.

Let $M$ be a spacetime which splits smoothly as $M = \R{} \times
S$, being its metric $g$ at each $z = (t,x) \in M$:
\begin{equation}\label{e6}
g((\tau,\xi),(\tau,\xi)) = - \beta (z) \tau ^{2} + 2 <\delta
(z),\xi > \tau + <\alpha _{z}(\xi ),\xi >, \end{equation} for all
$(\tau,\xi )\in T_{z}M \equiv \R{} \times T_{x}S$. Here, $\beta$
is a positive function on $M$, $<\cdot ,\cdot>$ denotes a fixed
auxiliary  Riemannian metric on $S$ (with associated norm
$\parallel \cdot
\parallel$ and distance {\it $d(\cdot ,\cdot )$}), $\alpha
_{z}$ is a symmetric positive operator on $T_{x}S$ and $\delta (z)
\in T_{x}S$, all varying smoothly with $z$. The 
 minimum eigenvalue of $\alpha _{z}$ will be denoted
$\lambda _{min}(z)$; notice that the eigenvalues of $\alpha _{z}$
vary (a priori just) continuously with $z$.

No more generality would be obtained if we put $M=I\times S$ for
some interval $I\subset \R$, as a re-scaling of the projection
$t:M\rightarrow I$ would reduce this case to the former one.
Analogously, we will assume that $<\cdot ,\cdot >$ is {\em
complete} with no loss of generality (we are free to choose any
auxiliary Riemannian metric, and different choices would redefine
$\delta$ and $\alpha$). It is straightforward to check that, under
our assumptions, $g$ must be Lorentzian, and we assume that the
future time-orientation is defined by $\partial_t$.

Now, notice that an inextensible future-directed causal curve
$\gamma(s)=(t(s),x(s))$ can be reparameterized by the coordinate
$t$, and will cross all the slices $S_t=\{t\}\times S_0$ if and
only if the curve $t\rightarrow \bar x(t)=x(s(t))$ can be
continuously extended to any finite value of $t$. Thus the slices
$S_t$ will be Cauchy if the curves type $\bar x(t)$ which come
from a causal curve have finite $\langle\cdot,\cdot\rangle$-length
for finite values of $t$. Then, a computation shows \cite[Sect.
3]{Sa-bari}:

\newpage
\begin{pspicture}(-0.25,1.25)

\rput(0,0){\rnode{A}{ \psframebox{
 \begin{tabular}{c}
  Spacelike\\
  Cauchy hyp.
 \end{tabular}}}}

\rput(4.7,0){\rnode{B}{ \psframebox{
 \begin{tabular}{c}
  Smooth and orthogonal\\
  Cauchy Splitting\\
  $g=-\beta dt^2+g_t$
 \end{tabular}}}}

\rput(9.85,0){\rnode{C}{ \psframebox{
 \begin{tabular}{c}
  Smooth and orthog.\\
  Cauchy Splitting\\
  with bounded lapse $\beta$
 \end{tabular}}}}

\rput(0,-2.5){\rnode{D}{ \psframebox{
 \begin{tabular}{c}
  Topological\\
  Cauchy hyp.
 \end{tabular}}}}

\rput(4.7,-2.5){\rnode{E}{ \psframebox{
 \begin{tabular}{c}
  Topol. acausal\\
  Cauchy splitting
 \end{tabular}}}}

\rput(0,-5){\rnode{F}{ \psframebox{
 \begin{tabular}{c}
  Compactness\\
  properties of\\
  $\CC$ 
 \end{tabular}}}}

\rput(4.8,-5){\rnode{I}{ \psframebox{
 \begin{tabular}{c}
  $J(p,q)$  
  compact\\
  $+$ strong caus.
 \end{tabular}}}}

\rput(9.75,-5){\rnode{J}{ \psframebox{
 \begin{tabular}{c}
  $J(p,q)$ compact\\
  $+$ causality
 \end{tabular}}}}


\rput(4.8,-8){\rnode{G}{ \psframebox{
 \begin{tabular}{c}
  Causal boundary\\
  with no\\
  timelike points
 \end{tabular}}}}

\rput(9.95,-8){\rnode{H}{ \psframebox{
 \begin{tabular}{c}
  Conformal boundary\\
  with no timelike\\
  points ($^*$) 
 \end{tabular}}}}

\ncline{->}{B}{A} \ncline{->}{C}{B}

\ncarc{->}{A}{D} \ncarc{->}{D}{A} \naput{\cite{BeSa1}}

\ncarc{->}{B}{E} \ncarc{->}{E}{B} \naput{\cite{BeSa2}}

\ncline{->}{E}{D}

\ncline{->}{E}{C} \nbput{\cite{MuSa}}

\ncline{->}{D}{F} \nbput{\cite{Ge}}

\ncline{->}{F}{E} \naput{\cite{Ge}}


\ncarc{->}{I}{F} \naput{\cite{Ge, Se}}

\ncarc{->}{F}{I}
\naput{\cite{CB, Ge}}

\ncline{->}{G}{H} \ncline{->}{H}{G} \nbput{\cite{FHS}}
\naput{\cite{PeBatelles}}

\ncline{<->}{I}{G} \nbput{\cite{FHS, GKP}}

\ncarc{<-}{I}{J} \naput{\cite{BeSa4}} \ncarc{<-}{J}{I}

\end{pspicture}




\vspace{10cm}
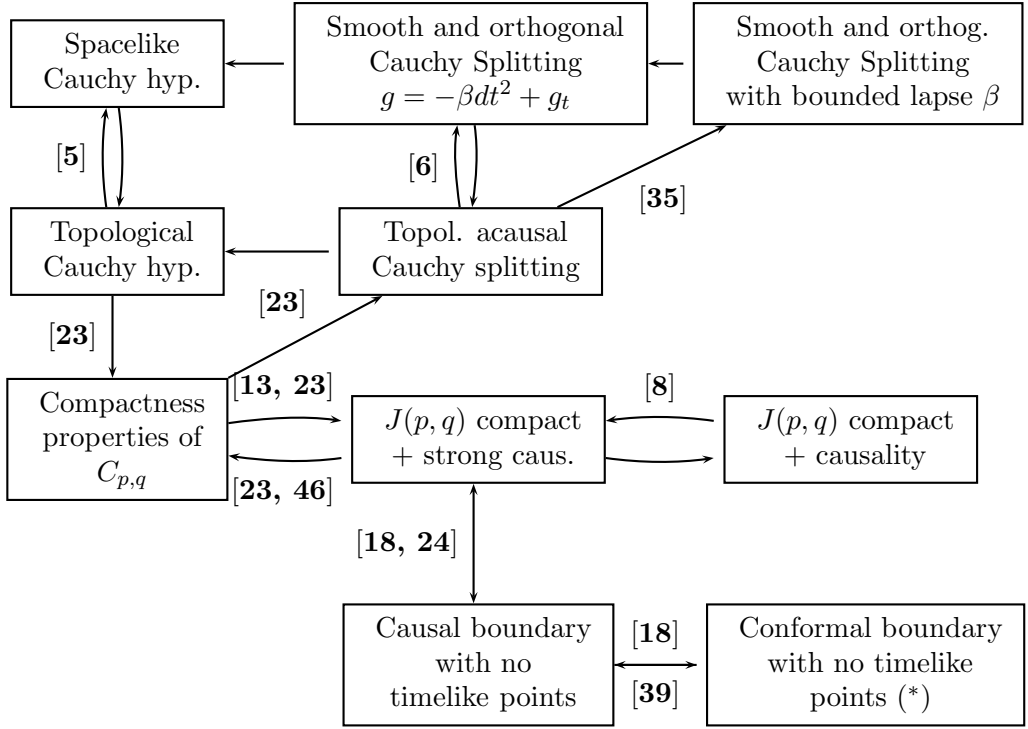
{\small \noindent {\em Figure.} Summary of both,  the classical
and revisited notions of global hyperbolicity:

In the first row, the characterizations of global hyp. involve
differentiability and metric properties. They imply trivially the
second row. The converses solve the so-called ``folk problems'' of
smoothability, first  pointed out in \cite{SW}.

The equivalence between the second and third rows relies on the
fundamental theorem by Geroch \cite{Ge}. He used a notion of
global hyp. based on the space of curves $\CC$ introduced by Leray
\cite{Le}. The  simplification of this definition in terms of
$J(p,q)$ in the third row, become widely convenient both,
technically (see, in general, Hawking and Ellis
\cite{HE}) 
and conceptually (the compactness of $J(p,q)$ can be understood as
the absence of naked singularities).

The main problem to prove the equivalences between the third and
the fourth rows, was to find consistent definitions for both, the
causal \cite{GKP} and the conformal \cite{PeBatelles} boundaries.
The drawn equivalence with a property of the causal boundary (as
defined in \cite{FHS}), is completely general. Nevertheless, the
conformal boundary can be defined only in some cases.

($^*$) The equivalence for the conformal boundary holds when a
chronologically complete open conformal embedding with $C^1$
boundary in a strongly causal spacetime (according to the
definitions in \cite{FHS}) exists. }

\newpage

\begin{theo}
\label{t6} For each positive integer $n$, put $M[n] = [-n,n]\times
S \subset M$, and assume that there exists a smooth function
$F_{n}$ on $S$ for each $n$ such that:
\begin{enumerate} \item the following inequality holds for all $(t,x) \in M[n]$:
$$
\frac{\parallel \delta \parallel +
(\lambda_{min} \beta + \parallel \delta \parallel
^{2})^{1/2}}{\lambda _{min}}(t,x) \leq F_{n}(x), $$

\item the conformal metric $<\cdot ,\cdot >_{n} = <\cdot ,
\cdot>/F_{n}^{2} \, $ on $S$ is also complete.
\end{enumerate}
Then each slice $S_{t} = \{ t\} \times S$ is a Cauchy surface.

In particular, this happens if  the following constants $M_n$
satisfy:

\[
M_{n} = {\rm Sup}\{ \frac{\parallel \delta (z)
\parallel}{\lambda_{min}(z) \, d_{0}(x)},
\sqrt{\frac{\beta(z)}{\lambda_{min}(z) \, d_{0}^{2}(x)}}:
z=(t,x)\in M[n], d_{0}(x)>1 \} < \infty
\]
where $d_{0}: S \rightarrow \R{}$ is the $<\cdot ,
\cdot>$--distance function to some (and then any) fixed point
$x_{0}\in S_0$.
\end{theo}

\begin{remark} (1) Of course, $\lim_{n}M_n=\infty$ is allowed.

(2) The hypothesis $d_0(x)>1$ in the definition of $M_n$ is
imposed because the relevant behavior of the metric elements
occurs at infinity. Equally, one can assume $d_0(x)>C$ for some
convenient constant $C$. Notice that, at any case, the
completeness of $\langle \cdot, \cdot \rangle$ implies the
compactness of the closed balls.

(3) In particular, if $S$ is compact then all the slices
$\{t\}\times S$ are Cauchy with no further assumption.
\end{remark}
It is not difficult to sharpen these estimates by using  subtler
bounds (say, bounding the elements of the splitting by suitable
radial functions with finite integral, so that $\bar x(t)$ will
not be divergent). We will not go into this general possibility,
but we will see next that a  clean characterization holds for
stationary spacetimes.

\section{Global hyperbolicity in standard stationary spacetimes}\label{s7}

Among the spacetimes which admit an expression as in (\ref{e6}),
appear the standard stationary ones. In fact, they are defined by
such a splitting when all the elements $\beta, \delta, \alpha$ are
independent of the coordinate $t$, i.e., the timelike vector field
$\partial_t$ is Killing\footnote{Recall that any distinguishing
spacetime with a complete timelike Killing vector field $K$ can be
written as a standard stationary \cite{JaSan08} (and, then,
$K\equiv \partial_t$). The proof of this result is also a
consequence of the solution of the folk problems of
smoothability.}. In this case, we can define a  metric $g_0 = \
<\alpha(\cdot),\cdot>$ on $S$, which is independent of $t$ and may
be incomplete. That is, for a standard stationary spacetime the
metric (\ref{e6}) turns out into:
\begin{equation}\label{e6b}
g((\tau,\xi),(\tau,\xi)) = - \beta (x) \tau ^{2} + 2 g_0(\delta
(x),\xi) \tau + g_0(\xi ,\xi),
\end{equation} for all $(\tau,\xi
)\in T_{z}M \equiv \R{} \times T_{x}S$.
 If, moreover, $\delta \equiv 0$ then $M$
is called standard static; that is, in this case $g= -\beta dt^{2}
+ g_0$, with natural identifications.

Theorem \ref{t6} yields directly the following consequence.
\begin{coro}
\label{coro 3.4} Assume that $(M,g)$ is standard  stationary,
according to (\ref{e6b}). If $g_0$ is complete and
\begin{equation}
\label{3.4a} {\rm Sup}\{ \frac{\parallel \delta
\parallel}{d_{0}}(x),
\frac{\sqrt{\beta}}{d_{0}}(x): x\in S, d_{0}(x)>1 \} < \infty,
\end{equation}
then the spacetime is globally hyperbolic and all the slices $S_t$
are Cauchy hypersurfaces.

In particular, (\ref{3.4a}) holds if there exist constants $a, b,
c, d$ such that, for all $x$ outside some compact subset:
$$
\parallel \delta (x) \parallel < a d_{0}(x) + b , \; \; {\rm and} \; \;
 \sqrt{\beta }(x) < c d_{0}(x) + d .
$$
\end{coro}
Notice that these results yield a rough estimate on the different
elements in the standard splitting so that the slices $S_t$ will
be globally hyperbolic. However, it is clear that these estimates
are only sufficient conditions. For example, in the static case
$g=-\beta dt^2 + g_0$, Corollary \ref{coro 3.4} imposes that $g_0$
is complete and $\beta$ grows at most quadratically with the
distance at infinity. Notice that both conditions are sufficient
to ensure that $g_0/\beta$ is complete. However, as Causality is
conformally invariant, one can study directly the global
hyperbolicity of $g^* = -dt^2 + g_0/\beta$. Then, Corollary
\ref{coro 3.4} applied to $g^*$ shows that the completeness of
$g_0/\beta$ ensure that the slices $\{t\}\times S$ are Cauchy (for
$g^*$ and, thus, for $g$). Moreover,  a simple computation
\cite[Th. 3.67]{BEE} shows that the converse holds and,
additionally, that $g$ is globally hyperbolic only if the
hypersurfaces $S_t$ are Cauchy.

One can generalize this sharpened results to standard stationary
spacetimes as follows. We will characterize both, when the slices
$S_t$ are Cauchy hypersurfaces and also when $M$ is globally
hyperbolic, even if the slices $S_t$ are not Cauchy. This
characterization is obtained in terms of the {\em Fermat metric}
associated to the standard stationary splitting. This is a Finsler
metric on $S$ of Randers type. The characterization, as well as
many consequences, is developed in detail in \cite{CJS}. Here, we
only describe it very briefly.

Given the standard stationary splitting (\ref{e6b}) the associated
Fermat metric on $S$ is defined as:
\begin{equation}\label{fermat}
F(v)=\frac{1}{\beta}g_0(v,\delta)+\sqrt{\frac{1}{\beta}g_0(v,v)+\frac{1}{\beta^2}g_0(v,\delta)^2},
\quad \forall v\in T_pS, p\in S. \end{equation} This is a type of
non-reversible Finsler metric\footnote{Non-reversible means
$F(v)\neq F(-v)$, in general. We remind that a Finsler metric
yields a non-symmetric norm with strongly convex balls in each
tangent space $T_xS$ smoothly varying with $x\in S$, see for
example \cite{BCS01}.}, and it induces a (non-necessarily
symmetric) distance $d_F$ on $S$. Such a $d_F$ also induces
forward and backward closed balls, defined respectively as:
$$\bar B^+(x,r)=\{y\in M : d_F(x,y)\leq r\}, \quad  \bar B^-(x,r)=\{y\in M : d_F(y,x)\leq r\}.$$
Analogously, one has forward and backward Cauchy sequences, and
forward and backward  completeness. We can also define the
symmetrized distance $d_s(x,y)=(d_F(x,y)+d_F(y,x))/2$ and the
corresponding closed symmetrized balls $\bar B_s(x,r)$. Then, one
has \cite[Sect. 4]{CJS}:

\begin{theorem}\label{t7}
For a standard stationary  spacetime $M$, the following properties
are equivalent:
\begin{enumerate}
\item  $M$ is globally hyperbolic.

\item The closed symmetrized balls $\bar B_s(x,r)$ of the Fermat
metric $F$ in (\ref{fermat}) associated to  one (and then to any)
standard stationary splitting of $M$ are compact.
\end{enumerate}
Moreover, the slices associated to a standard stationary splitting
are Cauchy hypersurfaces if and only if the Fermat metric
associated to that splitting is both, forward and backward
complete.
\end{theorem}
\begin{remark}
For a standard static spacetime ($\delta \equiv 0$)  the Fermat
metric is just the reversible Finsler metric $F=\sqrt{g_0/\beta}$,
and its associated distance is equal to the distance associated to
the Riemannian metric $g_0/\beta$. So, one recovers the
equivalence between:

(a) global hyperbolicity,

(b) completeness of $g_0/\beta$, and

(c) each $S_t$ is Cauchy.

\noindent However, in the general stationary case the condition
(b) splits into:

(b1) $\bar B_s(x,r)$ are compact, and

(b2) $d_F$ is forward and backward complete.

\noindent One has only the equivalences (a)$\Leftrightarrow$ (b1),
(b2)$\Leftrightarrow$ (c), plus the trivial implications (a)
$\Leftarrow$ (c), (b1) $\Leftarrow$ (b2).

Finally, notice that any spacelike Cauchy hypersurface $S'$ of a
standard stationary space induces a standard stationary splitting
such that $S'$ is one of the levels $t=$ constant. So, all
spacelike Cauchy hypersurfaces are characterized in terms of a
Fermat metric.
\end{remark}
Further properties of the Fermat metric in connection with both,
the causal boundaries and other boundaries for Riemannian and
Finsler manifolds, are being developed in \cite{FHSst}.

\section*{Acknowledgements}

The reading and comments by A.N. Bernal are warmly acknowledged. 
This research has been partially supported by the Regional J.
Andaluc\'{i}a Grant FQM 04496  and the national Spanish MEC Grant
MTM2007-60731, with FEDER funds.

\bibliographystyle{amsalpha}

\end{document}